# A damage detection method of plate structure using fan-shaped sensor clusters[*]

HAN Yue[1], MA Chenning[1], LIU Jinxia[1, *], ZHOU Zixian[1], YAN Shouguo[2], CUI Zhiwen[1, 2, 3, *]

1. College of Physics, Jilin University, Changchun 130021, China
2. State Key Laboratory of Acoustics and Marine Information, Chinese Academy of Sciences, Beijing 100190, China
3. Chongqing Research Institute of Jilin University, Chongqing 401120, China

**Abstract**

Plate structures are widely used in large-scale engineering fields such as aerospace, hull manufacturing, and construction. However, the plate structure is easily damaged during long-term service or when it is impacted by foreign objects. Such a damage may lead to serious safety accidents. Beamforming and L-shaped sensor cluster (LSSC) localization method can be used to locate the damage on the plate. However, when using beamforming or LSSC localization method to locate the damages on plate-like structures, there exists blind area. In this paper, by combining the beamforming and LSSC localization method, a fan-shaped sensor cluster is proposed through arranging five sensors in a fan shape, which can effectively reduce the blind areas. The positions of damages on the plate can be accurately detected by using two groups of fan-shaped sensor clusters and one sensor for transmitting the excitation signal. The feasibility of the fan-shaped sensor cluster localization method is verified through numerical simulations and experiments, and the results are compared with those obtained by using the T-shaped sensor cluster. The results show that the fan-shaped sensor cluster can more accurately identify the damages at different positions. Both simulation and experimental results indicate that the

## 1. Introduction

Plate structure is widely used in aerospace, hull manufacturing, construction and other large structures. When the structure is damaged due to long-term use, such as corrosion, leakage or foreign object impact, the stability of the structure will be affected. If the damaged location is not identified and repaired in a timely manner, it may lead to economic losses and even casualties [1]. Nondestructive testing (NDT) is a technology that can inspect or detect structural defects without damaging the structure [2,3]. It utilizes physical phenomena such as sound, heat, light, electricity, and magnetism to detect damage, thereby determining the location and extent of the damage [4–6]. Among various NDT methods, the detection technology based on Lamb waves has great application potential in the nondestructive testing of large plate structures [7–9]. Lamb waves are characterized by long propagation distances and high sensitivity to damage, making them suitable for damage localization in large structures [10–12]. When conducting detection using Lamb waves, sensors are typically placed on the surface of plate-like materials to actively generate excitation signals. Meanwhile, additional sensors are placed at other locations to receive the Lamb wave signals. Subsequent processing and analysis of the collected signals allow damage detection in the plate. [13,14].

With the rapid development of Lamb wave-based detection technology [15], various array signal processing methods are also developing rapidly. Beamforming has been widely used in the field of NDT in recent years due to its excellent performance in directional signal transmission and reception [16]. By performing operations such as phase shifting and superposition on the signals received by the array, it enables the superimposed signals to form a strong beam in a specific direction, focusing on receiving signals from the target direction while reducing interference from signals in other directions, thereby locating the damage position. For this technology, sensors are arranged in various ways, such as linear array, circular array, star array [17–19]. However, linear arrays are widely adopted due to their ease of use [20]. He et al. [21] employed a uniform linear array (ULA) and utilized near-field beamforming technology to predict the position of acoustic emission sources. Li et al. [22] used ULA to locate the continuous leakage source of $CO_2$ gas. In order to avoid the influence of vibration and noise, Zhang et al. [23] proposed the AF-MUSIC method, which can predict the impact source of composite plates under vibration conditions. Jung et al. [24] proposed a non-uniform distance linear array to avoid spatial aliasing and improve the robustness of localization. Beamforming has high resolution in the vertical direction, so it can accurately predict the source located in the vertical direction of the array, but there is an inevitable blind area in the horizontal direction. In order to eliminate the blind area of localization, two-dimensional structure are commonly adopted. Wang et al. [25] proposed a cross array, which can detect damage in all directions. Zhong et al. [26] added one sensor above and one sensor below the linear array to avoid horizontal blind area. Yu et al. [27] proposed a rectangular array that can perform omnidirectional detection. Although the above methods can reduce the blind area, they need to use a large number of sensors, which significantly increases the cost.

The L-shaped sensor cluster (LSSC) localization method proposed by Kundu et al.[28,29] is an effective method to predict the direction of arrival (DOA) based on the time difference of arrival (TDOA) of the received signals between sensors. This method does not require prior knowledge of material properties and complex data processing methods, and has great potential in the field of NDT[30]. Yin et al. improved LSSC[31] and proposed a Z-shaped sensor to improve the localization accuracy, the cross-shaped sensor cluster[32] is proposed and applied to the location of microcracks, which is efficient and accurate. Sen et al. proposed a square sensor cluster composed of four sensors [33], and each square sensor cluster can get four DOAs to further reduce the error. Zhou et al. [34] realized the active detection of composite plate damage by adding a sensor transmitting excitation signal on the basis of two groups of LSSC. LSSC location method is a fast and simple location method, but due to the nonlinear relationship between TDOA and DOA, it fails to accurately locate the damage in the vertical direction of the array.

Localization blind area has always been a problem worthy of attention in the field of NDT. In order to eliminate the blind area, Gao et al. [35–37] proposed the T-shaped sensor cluster localization method and the nonequidistant T-shaped sensor cluster localization method. By combining beamforming with the LSSC method, these approaches effectively reduce localization blind zones and improve both the accuracy and stability of localization.

In order to further reduce the blind area of damage location, based on the above work, a damage location method with fan-shaped sensor clusters is proposed in this paper. Through two groups of sector sensor clusters and one sensor transmitting excitation signal, the beamforming and LSSC location methods are combined to enhance the accuracy of localization results. The simulation and experiment are carried out, and the damage location results using this method are compared with those using T-shaped sensor cluster. The simulation and experimental results show that the proposed method can enhance the accuracy of localization results. This method can provide a new approach for damage detection of large structures.

## 2. Theory for damage localization

**2.1 Theory of damage location method for fan-shaped sensor cluster**

The fan-shaped sensor cluster structure proposed in this paper consists of five sensors. Among them, three sensors are arranged in an equidistant linear array, while the remaining two sensors form a set of symmetric L-shaped structures with the sensor at the center of the linear array as the vertex. The L-shaped sensor cluster is an isosceles right triangle structure, and the side length of the right-angle side is the same as the distance of the linear array. Xue et al. [38] confirmed through formula derivation and experimental verification that when the source is in the 45 ° direction relative to the LSSC, the LSSC localization method can accurately locate the source. Therefore, for the fan-shaped sensor cluster, the angle between the right-angle sides of the isosceles right triangle and the linear array is set to 45 °, and the sector sensor cluster is arranged as shown in the Fig.1.

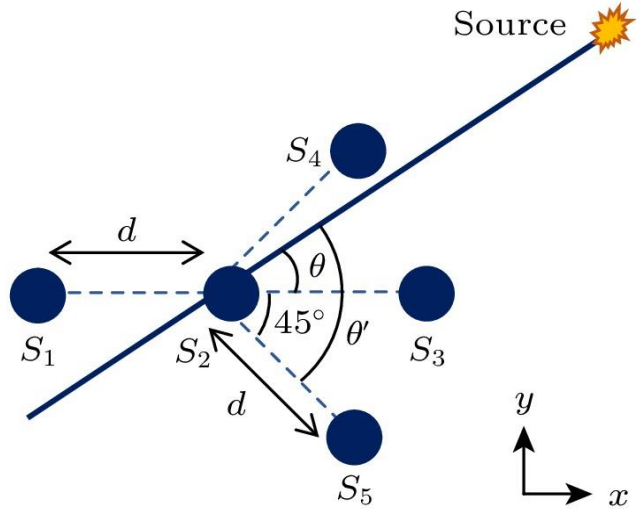

**Figure 1.** Schematic diagram of a fan-shaped sensor cluster.

When the damage has been formed, the stress wave is no longer generated at the damage location. At this time, the detection signal can be actively excited by an additional excitation source, and the damaged-scattered wave generated by the interaction between the excitation signal and the damage can be used for subsequent processing. Therefore, the scattered wave is regarded as a secondary sound source. Since only the far-field case is considered, i.e.,

$$r > \frac{2l^2}{\lambda} \tag{1}$$

Where $r$ is the distance between the damage and the array reference point, $l$ is the array length, and $\lambda$ is the signal wavelength. When the fan-shaped sensor cluster is used for damage location, the signal received by the linear array ($S_1S_2S_3$) is initially predicted by beamforming. After delaying and summing the signals received by the linear array along various directions, an angle-dependent energy function $E(\theta)$ can be obtained as,

$$E(\theta) = \int \left( \frac{1}{3} \sum_{i=1}^{3} \beta_i S_i'(t,\theta) \right)^2 dt \tag{2}$$

Where $S_i'(t)$ is the signal after delay processing, and $\beta_i$ is the weight coefficient independent of the signal. For the direction of the damage, the array signal is delayed and summed, and the output energy is maximum due to phase alignment. Draw a graph with energy as the ordinate and angle as the abscissa, and normalize the ordinate according to the maximum energy value. The angle corresponding to the maximum energy value is the preliminary prediction angle, as shown in the Fig. 2. If the angle corresponding to the maximum energy is not near 0 ° or 180 ° and there is an obvious main lobe, as shown in the Fig. 2(a), indicating that the damage location is not in the blind area of the linear array for damage location, this angle is the final predicted angle. On the contrary, if the angle corresponding to the maximum energy is near 0 ° or 180 °, and there is a flat area where it is difficult to distinguish the specific angle, as shown in the Fig. 2(b), it indicates that the damage is located in the blind area of the linear array, and the LSSC location method is needed to predict the received signal ($S_4S_2S_5$).

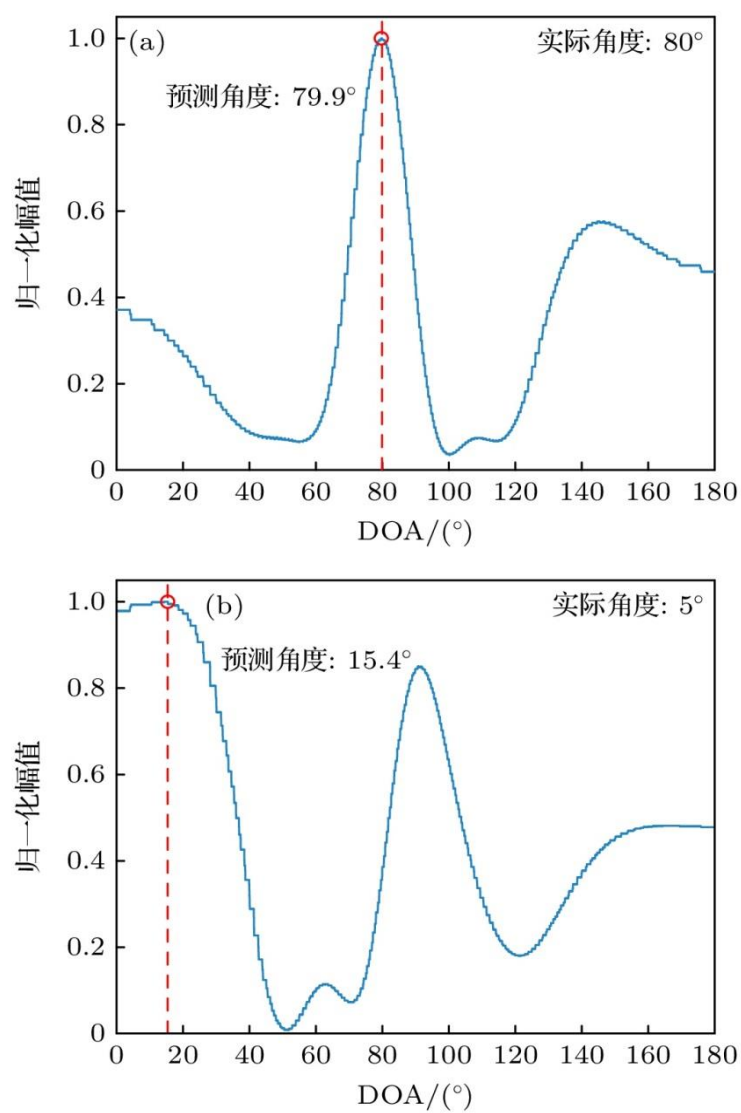

**Figure 2.** Schematic diagram of initial forecast results: (a) Prediction of an actual DOA of 80 °; (b) the prediction of an actual DOA of 5 °.

The secondary prediction needs to use the TDOA indirectly received by the sensor, which can be obtained by cross-correlation. Using the following formula, the angle $\theta'$ between the line (damage-S2) and the line (S2-S5) can be derived,

$$\Delta t_{42} = \frac{d \sin \theta'}{c(\theta')} \tag{3}$$

$$\Delta t_{52} = \frac{d \cos \theta'}{c(\theta')} \tag{4}$$

$$\theta' = \tan^{-1} \frac{\Delta t_{42}}{\Delta t_{52}} = \tan^{-1} \frac{y_0 - y_2}{x_0 - x_2} \tag{5}$$

Where $\Delta t_{ij}$ represents the TDOA between different sensors, $(x_0, y_0)$ represents the damage coordinate, $(x_2, y_2)$ represents the coordinate of the sensor $S_2$, and $c(\theta')$ represents the wave velocity in the $\theta'$ direction.

After the secondary prediction, the final damage prediction angle $\theta$ can be expressed as

$$\theta = |\theta' - 45°| \tag{6}$$

Therefore, when this method is applied to locate damage in a plate, two sets of fan-shaped sensor clusters can be placed on the plate. The two obtained predicted angles are then combined with the sensor positions to predict the damage location.

**2.2 Damage features**

When this method is used to detect the damage of panel structure, the waveform and TDOA of the damage-scattered signals received by different sensors are needed. Because the damage size is often small, the intensity of the damage scattering wave generated by the interaction between the damage and the excitation signal is much lower than that of the original excitation signal and the boundary reflection signal. Therefore, the difference signal obtained by subtracting the signal received in the damaged structure from the baseline signal received in the healthy structure can make the damage-scattered signals more clearly visible, which is easier for subsequent data processing.

**2.3 Error in damage localization**

When the fan-shaped sensor cluster localization method is used to detect the damage, the error between the predicted damage location and the actual damage location is expressed by the Euclidean distance between them, and the specific error calculation formula can be expressed as,

$$error = \sqrt{(y_p - y_a)^2 + (x_p - x_a)^2} \tag{7}$$

Where $(x_p, y_p)$ is the predicted damage coordinate and $(x_a, y_a)$ is the actual damage coordinate.

## 3. Numerical investigation and demonstration

Due to the convenience of the simulation software, the finite element simulation software COMSOL Multiphysics was used to verify the feasibility of the fan-shaped sensor cluster localization method before the experiment.

**3.1 Construction of simulation model**

A three-dimensional model was constructed using COMSOL Multiphysics, featuring an aluminum plate with dimensions of 500 mm (length) × 500 mm (width) × 2 mm (thickness). The material parameters are shown in Tab.1. According to the material properties, the phase velocity and group velocity dispersion curves of Lamb waves propagating in the aluminum plate are calculated, as shown in Fig.3 and Fig.4.

**Table 1.** Material parameters of aluminum plate.

| Material property | Numerical value |
|---|---|
| Density $\rho$/ (kg·m$^{-3}$) | 2700 |
| Poisson's ratio $\sigma$ | 0.33 |
| Young's modulus $E$/(GPa) | 70 |

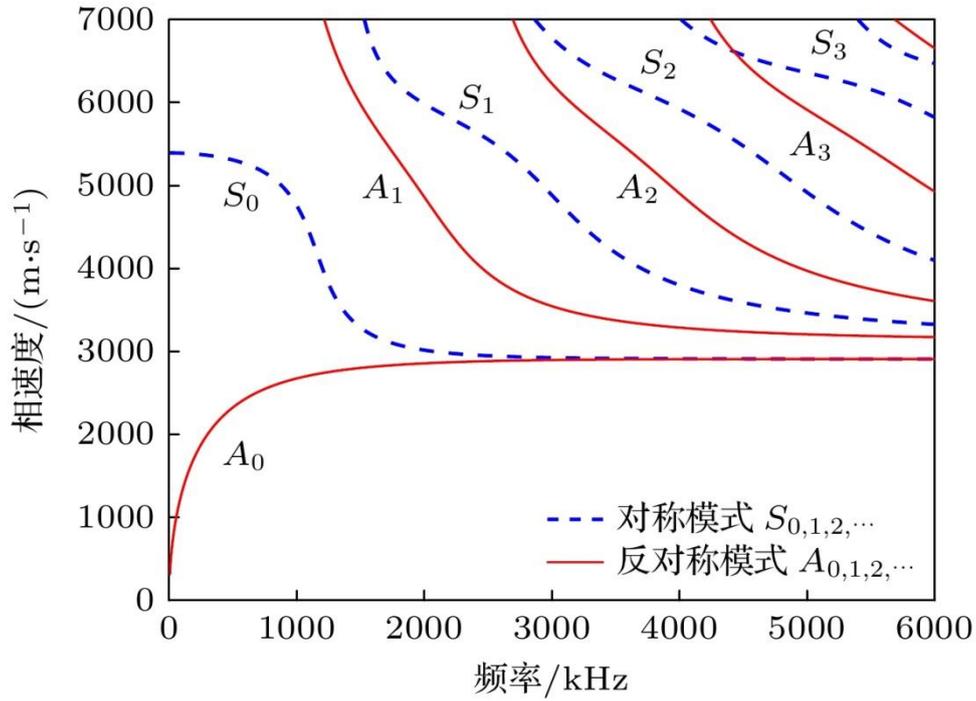

**Figure 3.** Phase velocity dispersion curves of an aluminum plate with a thickness of 2 mm.

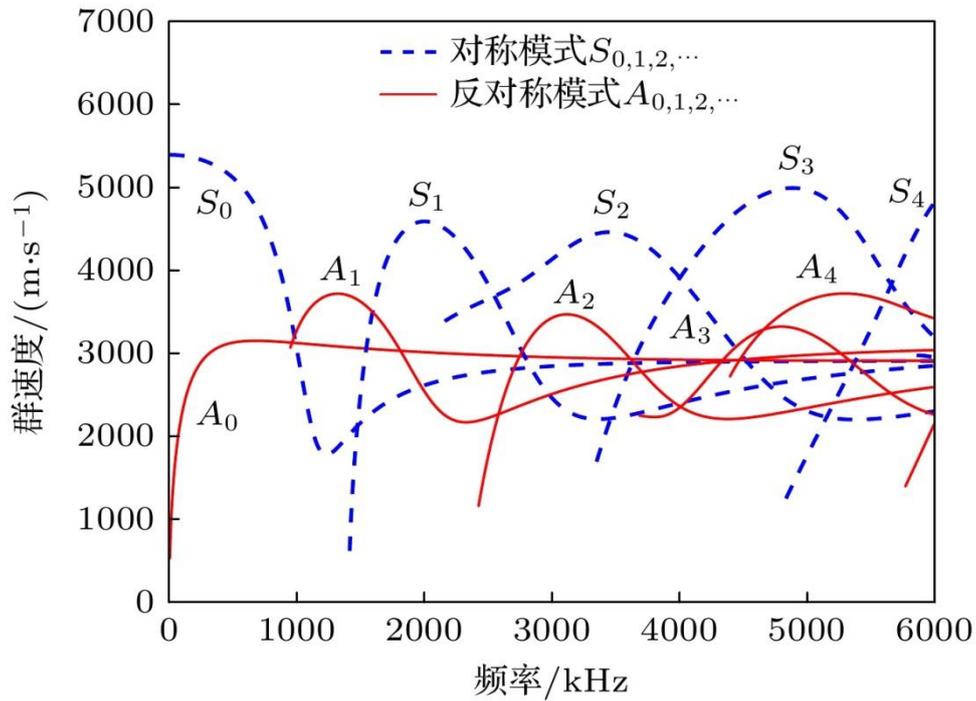

**Figure 4.** Group velocity dispersion curves of an aluminum plate with a thickness of 2 mm.

It can be seen from the dispersion curve that when the center frequency of the excitation signal is low, the Lamb wave propagating in the plate has only two modes, $A_0$ and $S_0$. When the frequency is high, there are many higher order modes of Lamb waves propagating in a plate. Because the wavelength of the $A_0$ mode is smaller than that of the $S_0$, the $A_0$ mode wave is more sensitive to damage, and the $A_0$ mode wave only produces the scattered $A_0$ mode when it encounters damage during propagation. Therefore, in order to avoid the influence of other mode waves on the detection results, the $A_0$ mode wave is used to carry out the detection work. In the selection of excitation frequency, in order to make the $A_0$ modal wave response large and avoid the influence of other modal waves, a resonant sensor with a center frequency of 150 kHz is generally used to detect thin plates in reference engineering, and the excitation frequency is finally set to 150 kHz. Considering the detection effect and the convenience of data processing, the pulse narrow-band signal shall be selected to simulate the acoustic emission source signal, and the excitation signal shall be finally set as a 3-cycle sinusoidal signal modulated by the Hanning window, and the expression is as follows,

$$z(t) = \frac{1}{2}[1-\cos(\frac{2\pi ft}{3})]\sin(2\pi ft) \tag{8}$$

A point load is placed at the center of the plate to be used as an excitation signal. Two sets of fan-shaped sensor clusters were positioned 50 mm above and below this point load, respectively, with a sensor spacing of 10 mm. The stress information at the sensors is recorded as a received signal. The coordinate of the excitation source is (0.00, 0.00) mm, and the position coordinate of the sensor is shown in Tab. 2. The geometry of the model and the location of the sensors are shown in the Fig. 5. The mesh was generated using free tetrahedrons, and the mesh size was set to 2 mm to satisfy the Courant-Friedrichs-Lewy condition. The total solution time was 180 μs, and the fixed step size was 0.1 μs.

**Table 2.** Coordinates of sensors.

| | Coordinates of sensors /mm | | |
|---|---|---|---|
| $S_1$ | (–10.00, –50.00) | $S_6$ | (–10.00, 50.00) |
| $S_2$ | (0.00, –50.00) | $S_7$ | (0.00, 50.00) |
| $S_3$ | (10.00, –50.00) | $S_8$ | (10.00, 50.00) |
| $S_4$ | (7.07, –42.93) | $S_9$ | (–7.07, 57.07) |
| $S_5$ | (7.07, –57.07) | $S_{10}$ | (–7.07, 42.93) |

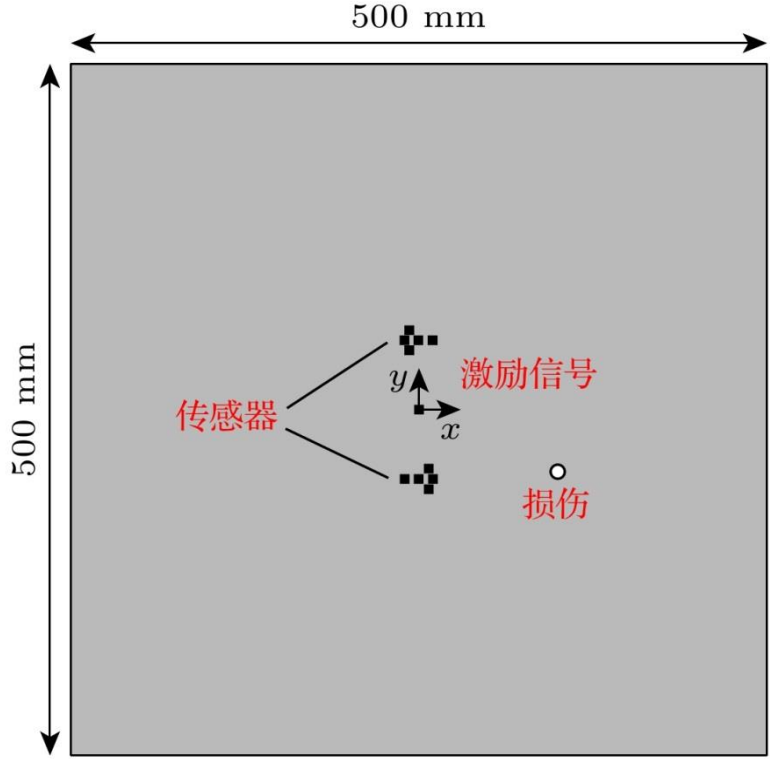

**Figure 5.** Schematic diagram of sensors layout in the simulation.

In order to verify the stability of this method, a total of eight damages at different positions were selected, with the center positions of the damages provided in Tab.3. Meanwhile, to facilitate the observation of damage-scattered signals, through-holes with a diameter of 10 mm were used to simulate the damages. When the damage is located at (100,-45), the wave field snapshot at the 70 μs is shown in the Fig.6. It can be seen that the elastic wave propagates with a circular wavefront, and damage-scattered waves generated by the interaction between the excitation signal and the damage are visible near the damage.

**Table 3.** Simulation localization results and errors.

|  | Actual damage coordinate/mm | T-shaped sensor cluster | | Fan-shaped sensor cluster | |
| --- | --- | --- | --- | --- | --- |
|  |  | Predicted damage coordinate/mm | Error /mm | Predicted damage coordinate/mm | Error /mm |
| D1 | (46.00, 123.00) | (45.93, 125.16) | 2.16 | (46.29, 125.73) | 2.75 |
| D2 | (−90.00, 20.00) | (−75.34, 11.45) | 16.97 | (−94.02, 24.28) | 5.87 |
| D3 | (130.00, 60.00) | (133.83, 70.07) | 10.77 | (133.48, 64.39) | 5.61 |
| D4 | (30.00, −100.00) | (34.30, −107.32) | 8.49 | (32.96, −105.08) | 5.88 |
| D5 | (−60.00, −60.00) | (−54.31, −63.35) | 9.27 | (−54.58, −56.65) | 6.37 |

|  | Actual damage coordinate/mm | T-shaped sensor cluster | | Fan-shaped sensor cluster | |
| --- | --- | --- | --- | --- | --- |
|  |  | Predicted damage coordinate/mm | Error /mm | Predicted damage coordinate/mm | Error /mm |
| D6 | (105.00, –10.00) | (93.94, –6.28) | 11.67 | (105.19, –6.87) | 3.14 |
| D7 | (–35.00, 115.00) | (–35.72, 118.03) | 3.11 | (–33.22, 113.27) | 2.48 |
| D8 | (100.00, –45.00) | (84.55, –33.09) | 19.51 | (97.29, –44.28) | 2.80 |

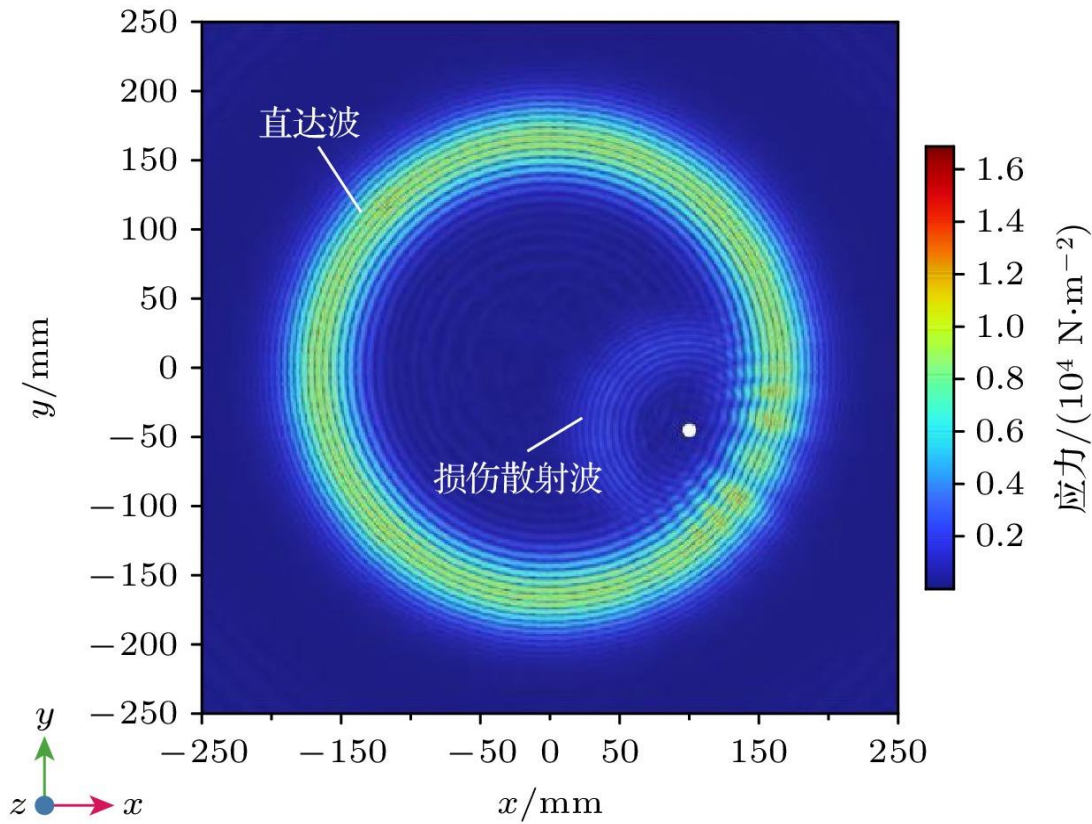

**Figure 6.** Wavefield snapshot at $t = 70$ μs (the damage is located at (100, –45)).

At this point, the comparison between the damaged signal received by $S_1$ of the lower sector sensor cluster and the baseline signal are shown as Fig.7. In the Fig.7(a), the black line represents the signal received when there is damage, and the red line is the baseline signal. When damage exists, the received signal can be divided into three obvious wave packets, the first wave packet is the excitation signal received by the $S_1$, the second wave packet is the scattering signal generated by the interaction between the excitation signal and the damage, and the third is the boundary-reflected wave. By performing a time-domain coherent subtraction between the signal received from the damaged structure and the baseline signal, a difference signal is obtained, which enables the effective separation of the damage-scattered wave, as shown in Fig.7(b).

Similarly, the damage-scattered signals of other sensors are also obtained by this method, and the damage-scattered signals are subsequently processed.

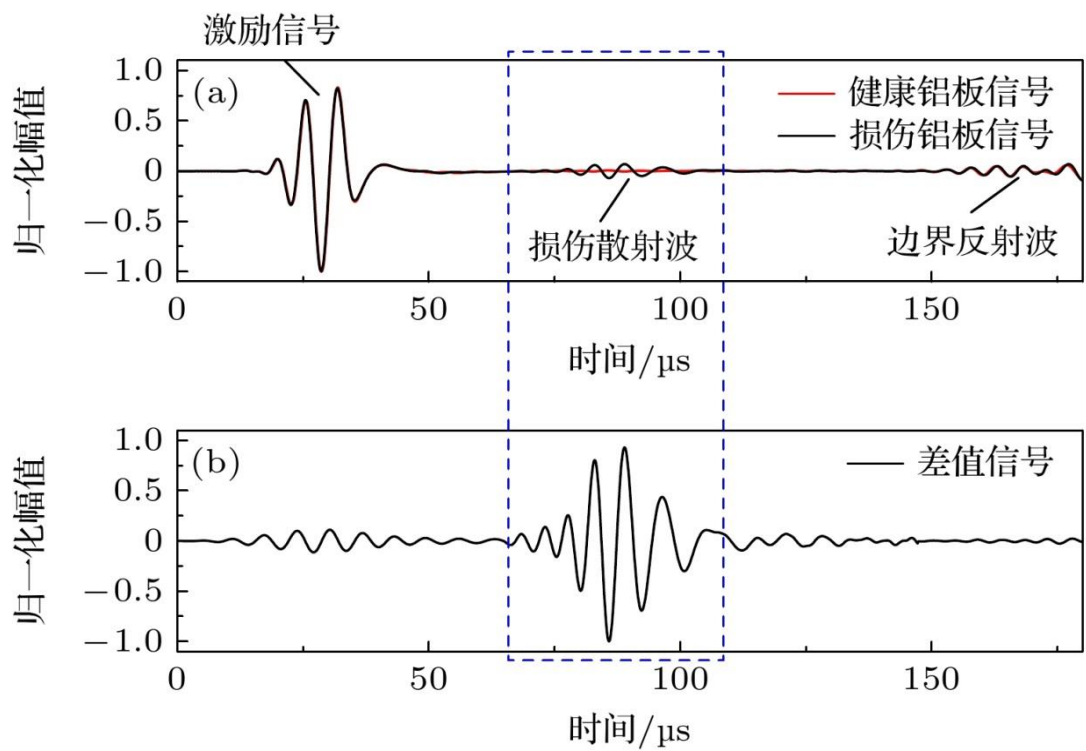

**Figure 7.** (a) Damaged signal and the healthy signal received by the $S_1$ when the damage is located at (100, –45); (b) the differential signal obtained by subtracting the healthy signal from the damaged signal received by the sensor $S_1$.

**3.2 Discussion of simulation result**s

The received damage-scattered signals were processed using the fan-shaped sensor cluster damage localization method, then the predicted damage location can be obtained. The prediction results and errors are shown in the Tab.3 and Fig.8(b). For comparison, the T-shaped sensor clusters are placed at the same location, and the T-shaped sensor cluster localization method is used to simulate the eight damage locations, and the prediction results and errors are shown in the Tab.3 and Fig.8(a).

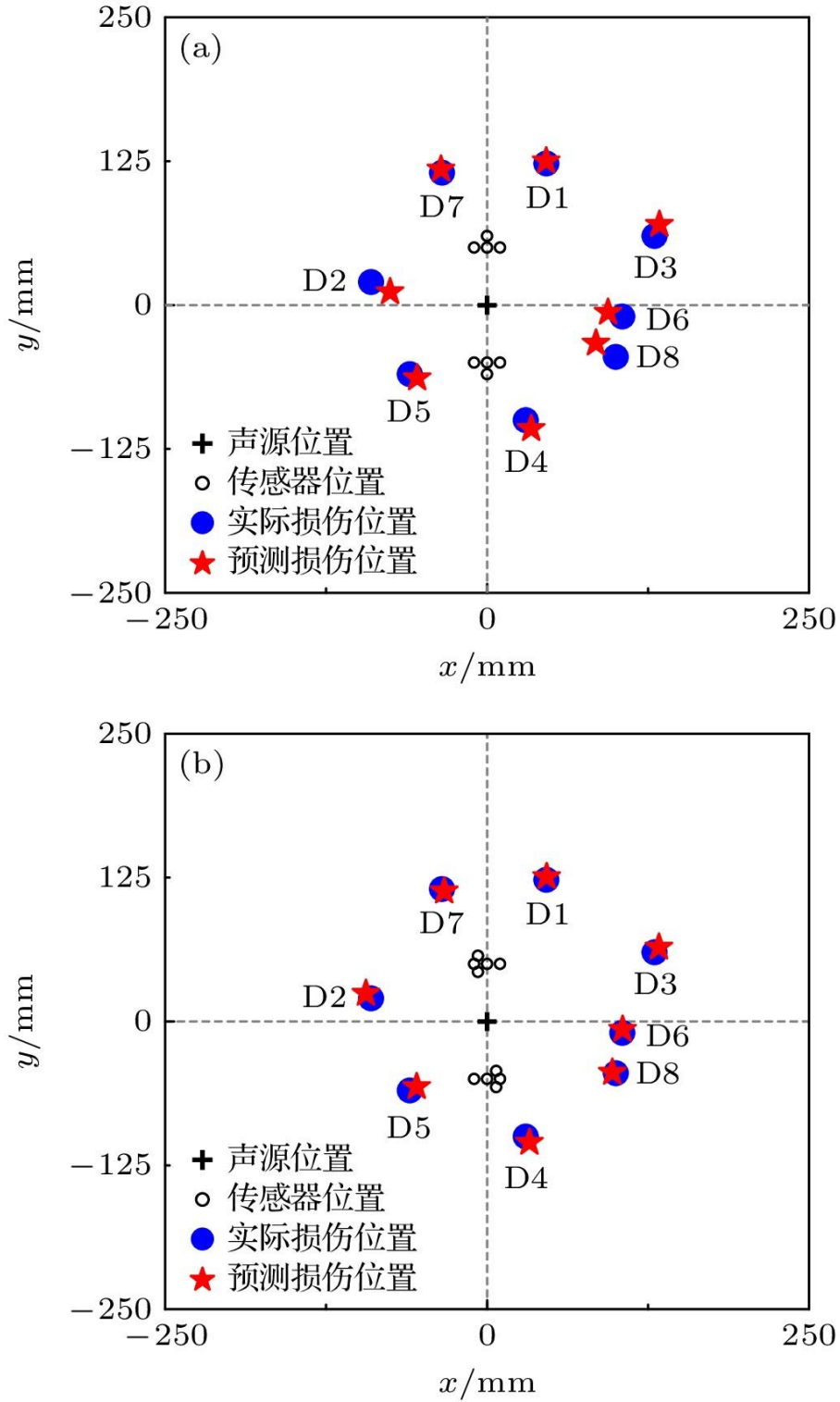

**Figure 8.** Schematic diagram of simulation localization results: (a) Localization results of T-shaped sensor cluster; (b) localization results of fan-shaped sensor cluster.

The simulation results show that the damage location method based on fan-shaped sensor clusters can accurately detect the damage located near the horizontal direction of the sensor array. In contrast, when using the T-shaped sensor cluster for damage detection, there exist certain damage localization blind area. For example, for the damages D2, D3, D5 and D8 which are close to the horizontal direction of the linear array. However, the error is smaller when use the fan-shaped sensor cluster for localization. For damages in other directions, the fan-shaped sensor cluster can also detect them accurately.

The fan-shaped sensor cluster localization method effectively improves the localization accuracy. Moreover, among the aforementioned damage cases, the maximum prediction error does not exceed 6.37 mm. The simulation results thus verify the feasibility of this method.

## 4. Experimental verification

In order to further verify the effectiveness of the damage location method of the fan-shaped sensor cluster, physical experiments were carried out on the aluminum plate under laboratory conditions. Two aluminum plates of the same material and size of 500 mm (length) × 500 mm (width) × 2 mm (thickness) were used in the experiment. One plate was artificially machined with a through hole of about 10 mm in diameter to simulate damage, and the other plate was non-damaged to receive baseline signals. OLYMPUS 5800 pulse signal transmitter, oscilloscope and AE144S sensor with resonant frequency of 150 kHz are used to form the experimental device, as shown in Fig. 9. The sampling frequency is 10 MHz.

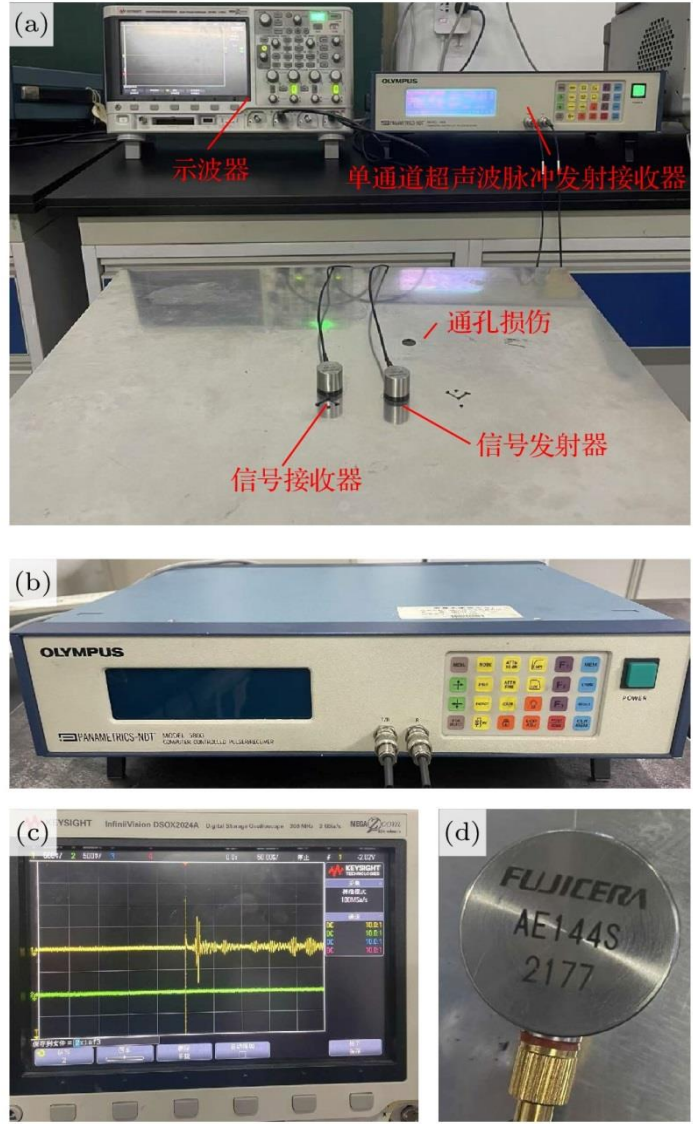

**Figure 9.** (a) Photo of the experimental setup; (b) OLYMPUS 5800 pulse signal transmitter; (c) the signal received by the oscilloscope in the experiment; (d) AE144S sensor.

The arrangement of the sensors is the same as that in the simulation. In the experiment, the damage at different positions is simulated by rotating the damaged aluminum plate and changing the relative position between the damage and the excitation source. A total of 8 damages at different locations are selected, and the coordinates of the damage center are the same as those in the simulation, as listed in Tab. 4. When the damage is located at (105, – 10), the damage signal received by the sensor $S_1$ and the baseline signal are shown in the Fig. 10(a), and the damage-scattered signal can be obtained by subtracting the signal received by the sensor with damage from the baseline signal, as shown in the Fig. 10(b). In the same way, the damage signals received by other sensors are also obtained by this method, and the damage-scattered signals are processed.

**Table 4.** Experimental localization results and errors.

| | Actual damage coordinate/mm | T-shaped sensor cluster | | Fan-shaped sensor cluster | |
|---|---|---|---|---|---|
| | | Predicted damage coordinate/mm | Error /mm | Predicted damage coordinate/mm | Error /mm |
| D1 | (46.00, 123.00) | (42.39, 115.11) | 8.68 | (43.22, 114.82) | 8.64 |
| D2 | (−90.00, 20.00) | (−106.88, 22.25) | 17.03 | (−96.76, 20.04) | 6.76 |
| D3 | (130.00, 60.00) | (141.36, 66.56) | 13.12 | (124.69, 56.48) | 6.37 |
| D4 | (30.00, −100.00) | (29.52, −97.61) | 2.44 | (31.51, −103.49) | 3.80 |
| D5 | (−60.00, −60.00) | (−65.45, −60.33) | 5.46 | (−65.40, −60.12) | 5.40 |
| D6 | (105.00, −10.00) | (94.63, −12.02) | 10.56 | (111.13, −14.78) | 7.77 |
| D7 | (−35.00, 115.00) | (−36.40, 121.49) | 6.64 | (−40.94, 119.23) | 7.29 |
| D8 | (100.00, −45.00) | (90.54, −33.69) | 14.74 | (106.07, −47.88) | 6.72 |

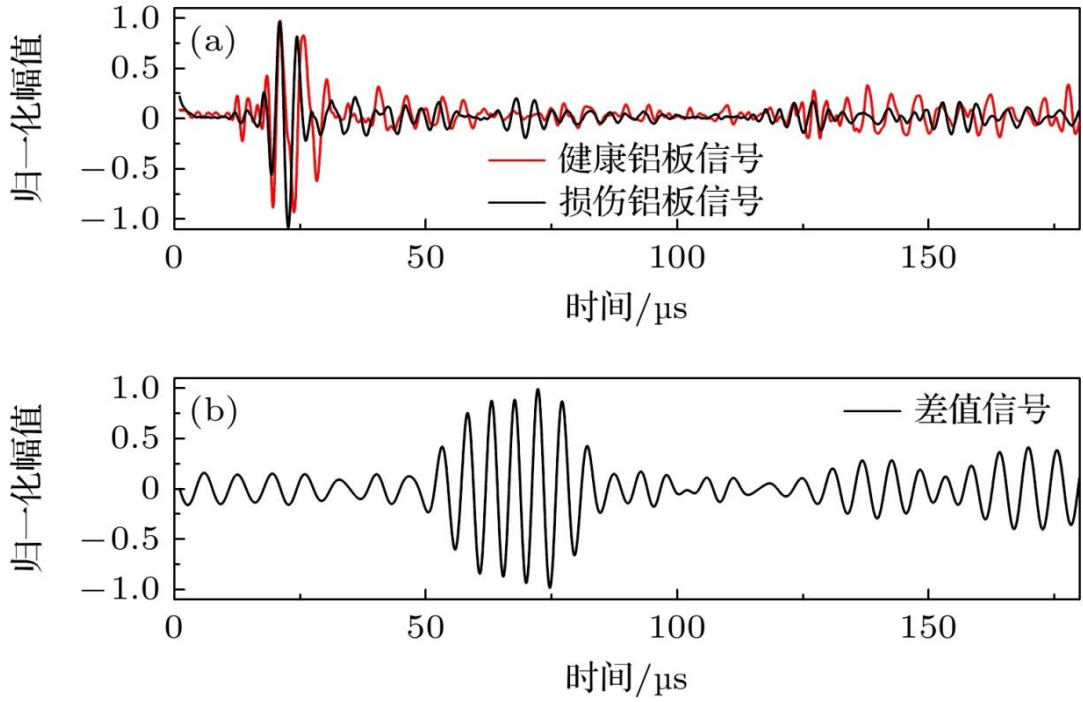

**Figure 10.** (a) Damaged aluminum signal and the healthy aluminum signal received by the $S_3$ of the fan-shaped sensor cluster on the lower side when the damage is located at (105, −10); (b) the differential signal obtained by subtracting the healthy signal from the damaged signal received by the sensor $S_3$ after filtering.

The detection results and errors using the fan-shaped sensor cluster are shown in the Tab.4 and Fig.11(b). For comparison, the T-shaped sensor cluster is also used to predict the damage at the same location in the experiment, and the prediction results and errors are shown in the Tab.4 and Fig.11(a).

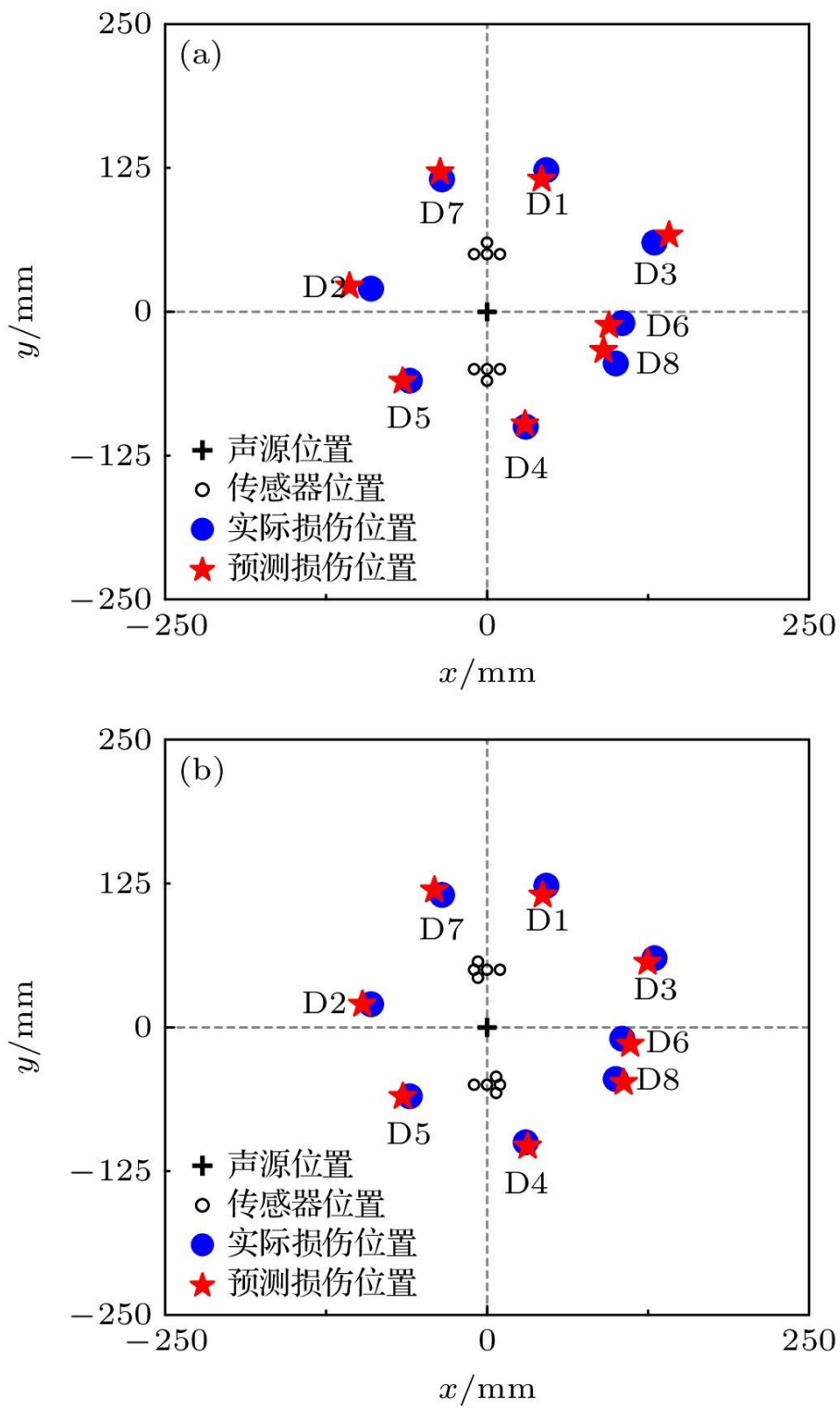

**Figure 11.** Schematic diagram of experimental localization results: (a) Localization results of T-shaped sensor cluster; (b) localization results of fan-shaped sensor cluster.

Similar to the above simulation results, the experimental results also show that the fan-shaped sensor cluster can accurately detect the damage on the plate. Moreover, compared with the T-shaped sensor cluster, the fan-shaped sensor cluster achieves higher prediction accuracy when detecting damages close to the horizontal direction of the sensor array. For example, for the damages D2, D3, D5 and D8 which are close to the horizontal direction of the linear array, the error is large when the T-shaped sensor cluster is used for detection. However, the error is smaller when use the fan-shaped sensor cluster for localization. For damages in other directions, the fan-shaped sensor cluster can also detect them accurately

The fan-shaped sensor cluster localization method enables more accurate damage localization. Moreover, among the aforementioned damage cases, the maximum prediction error does not exceed 8.64 mm. The experimental results thus confirm the feasibility of this method.

## 5. Conclusion

In this paper, a fan-shaped sensor cluster structure is innovatively proposed to improve damage localization accuracy, taking into account the advantages of beamforming and LSSC location methods, by combining the advantages of beamforming and LSSC localization methods. Through arranging five sensors in a fan-shaped structure, the blind area of damage location can be effectively reduced by using this geometric structure. The damage location in the plate can be accurately detected by using a detection system composed of two groups of fan-shaped sensor clusters and an acoustic source for exciting signals. In order to verify the effectiveness of this method, an aluminum plate is used for simulation and experimental verification, and the prediction results are compared with those of the T-shaped sensor cluster. Both simulation and experimental results show that the proposed method can effectively improve the accuracy of damage location.

The fan-shaped sensor cluster proposed in this study enables more accurate localization of damage positions. Compared with the T-shaped sensor cluster, the fan-shaped sensor cluster can accurately locate the damage near the horizontal direction of the array. And the method is easy to operate, because there is no damage location blind area, the fan-shaped sensor clusters can be placed without any constraints on orientation or position. There is no need to solve complex nonlinear equations for data processing. The damage location method of plate structure based on fan-shaped sensor cluster provides a new theoretical method for damage location, which is of great significance for the research of damage detection method of plate Lamb wave.